\documentclass[twocolumn,prl,showpacs,floatfix]{revtex4}

\usepackage{graphicx}
\usepackage{dcolumn}
\usepackage{bm}
\usepackage{amsmath}

\begin{document}

\title{Unified description of floppy and rigid rotating Wigner molecules\\
formed in quantum dots}

\author{Constantine Yannouleas}
\author{Uzi Landman}

\affiliation{School of Physics, Georgia Institute of Technology,
             Atlanta, Georgia 30332-0430}

\date{18 November 2003}

\begin{abstract}
Restoration of broken circular symmetry is used to explore the characteristics
of the ground states and the excitation spectra of rotating Wigner molecules 
(RWM's) formed in two-dimensional parabolic $N$-electron quantum dots. In high
magnetic fields, the RWM's are floppy rotors with the energies of the magic 
angular momentum $(L)$ states obeying $a L + b/L^{1/2}$. Under such fields the 
ground-state energies (referenced to the kinetic energy in the lowest Landau 
level) approach the electrostatic energy of $N$ point charges in the classical
equilibrium molecular configuration. At zero field and strong interelectron 
repulsion, the RWM's behave like quasiclassical rigid rotors whose energies 
vary as $L^2$. 
\end{abstract}

\pacs{73.21.La; 71.45.Gm; 71.45.Lr}

\maketitle

The belief that the physics of two-dimensional (2D) semiconductor quantum dots
(QD's) in high magnetic fields $(B)$ is described \cite{jai,haw} 
by composite-fermions, introduced \cite{jai2} originally for the bulk 
fractional quantum Hall effect (FQHE), has been recently challenged 
\cite{yl1} by a proposal that the fundamental physical entity in strongly 
correlated QD's is a {\it collectively rotating\/} Wigner (or electron) 
molecule (RWM or REM).
Indeed, it has been lately found that electrons in 2D QD's can
undergo in the strongly correlated regime a spontaneous phase transition akin 
to the Wigner crystallization in the bulk, forming \cite{yl2,yl3,yl4,mak} 
specific polygonal geometric structures that are called Wigner molecules
(WM's). These geometric structures break the 
rotational symmetry (for symmetry restoration, see below) and reflect 
the presence of many-body crystalline correlations that are revealed,
as illustrated previously, in the ``humps'' of the conditional 
probabilities \cite{yl4,mak} and of the broken-symmetry electron densities 
\cite{yl2,yl3}.

The majority of recent studies address the properties of {\it static\/} WM's 
(as a function of the QD parameters) \cite{reu,kai,sun,sza}. Here, we study 
the properties of the {\it rotating\/} WM.
We show through microscopic many-body investigations
that the RWM in high $B$ is a {\it floppy rotor\/}, while at large 
$R_W$ (and zero magnetic field) it transforms into a {\it rigid rotor\/}. 
[The Wigner parameter $R_W \equiv Q/\hbar \omega_0$, where $Q$ is the Coulomb 
interaction strength; $Q=e^2/(\kappa l_0)$, with 
$l_0 = \sqrt{\hbar/(m^* \omega_0)}$ being the spatial extent of the lowest 
single-electron wave function in the external parabolic confinement of 
frequency $\omega_0$, and $\kappa$ is the dielectric constant.] The ability to
capture the physics of the electrons in QD's in both the high magnetic field 
and the field-free regimes is an essential demonstration of the powerful 
unification offered by the RWM picture, a property not shared by other 
suggested approaches.

The {\it collective rotation\/} of the WM is inherent and natural to the 
molecular picture. In particular, we show that it is manifested in
characteristic energy-vs-angular-momentum relations for the excitation 
spectra and in specific limiting values for the ground-state energies.
These relations are important in the development of theories of
electrons in QD's under the influence of a magnetic-field and in the 
field-free case, and they can be employed as diagnostic tools for assessing 
the validity and applicability of alternative theoretical descriptions.

Description of the broad variation of the 
collective properties of electrons in 2D QD's
requires a highly flexible and accurate many-body method. In principle, exact 
diagonalization (EXD) could have been used; however, its computational 
limitations are a major obstacle. We 
demonstrate that the recently proposed \cite{yl3,yl1} two-step method of 
circular symmetry breaking at the unrestricted Hartree-Fock (UHF) level and of
subsequent symmetry restoration via post-Hartree-Fock projection techniques is
an accurate and computationally efficient approximation which provides a 
{\it unified\/} microscopic description of the emergent picture of RWM's in 
2D QD's.

The restoration of broken circular symmetry is necessary for a proper 
description of RWM's. Indeed, while the broken symmetry UHF solutions 
describe {\it static\/} WM's \cite{reu,kai,sun,sza}, it is the symmetry 
restoration step which describes the rotation of the WM's and underlies the 
differentiation between rigid and floppy (at high $B$) rotors.
 
{\it The two-step method.\/}
In general, the localized broken symmetry orbitals are determined 
numerically via a selfconsistent solution of the UHF equations \cite{yl3}. 
An efficient alternative, however, is to approximate these orbitals by
appropriate analytical expressions \cite{yl1,kai,sun,sza}. 
Since we focus here on the second step (restoration of the circular symmetry) 
and the case of high $B$, it will be sufficient to approximate the UHF orbitals
(first step of our procedure) by (parameter free) displaced Gaussian 
functions \cite{note5}; namely, for an electron localized at $Z_j$, we use 
the orbital 
\begin{equation}
u(z,Z_j) = \frac{1}{\sqrt{\pi} \lambda}
\exp \left( -\frac{|z-Z_j|^2}{2\lambda^2} - i\varphi(z,Z_j;B) \right),
\label{uhfo}
\end{equation}
with $z=x+iy$, $Z_j = X_j +i Y_j$, and $\lambda = \sqrt{\hbar /m^* \Omega}$;
$\Omega=\sqrt{\omega_0^2+\omega_c^2/4}$, where $\omega_c=eB/(m^*c)$ is the
cyclotron frequency. The phase guarantees gauge invariance in the presence of 
a perpendicular magnetic field and is given in the symmetric gauge by
$\varphi(z,Z_j;B) = (x Y_j - y X_j)/2 l_B^2$, with $l_B = \sqrt{\hbar c/ e B}$
being the magnetic length. We only consider the case of fully polarized 
electrons, which is appropriate at high $B$.

\begin{figure}[t]
\centering\includegraphics[width=7.5cm]{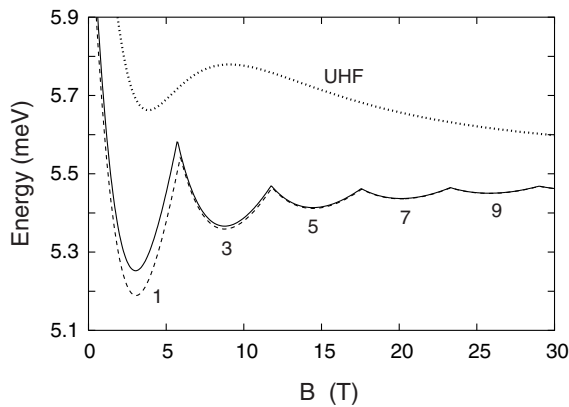}
\caption{
Total energies, $E-2 \hbar \Omega$, for the triplet state of N=2 electrons 
as a function of the magnetic field $B$. Solid line: PRJ (rotating WM); 
dashed line: exact; dotted line: UHF (static WM). 
The integers below the troughs of the PRJ and exact lines indicate the 
corresponding magic angular momenta. The parameters are: 
$m^*=0.067m_e$, $\hbar \omega_0 = 3$ meV, and $\kappa=12.9$. 
Here, as well as in the other figures, the Zeeman contribution is not 
included, but can be easily added.
}
\end{figure}
We take the $Z_j$'s to coincide with the equilibrium positions (forming nested
regular polygons \cite{note9}) of $N$ classical point 
charges inside an external parabolic confinement of
frequency $\omega_0$, and proceed to construct the UHF determinant 
$\Psi^{\text{UHF}} [z]$ out of the orbitals $u(z_i,Z_j)$'s, $i,j = 1,...,N$.
Correlated many-body states with good total amgular momenta $L$ can be 
extracted \cite{yl3} from the UHF determinant using projection operators; 
the projected energies are given by \cite{yl3}
\begin{equation}
E_{\text{PRJ}} (L) = \left. { \int_0^{2\pi} h(\gamma) e^{i \gamma L}
d\gamma } \right/ { \int_0^{2\pi} n(\gamma) e^{i \gamma L} d\gamma},
\label{eproj}
\end{equation}
with $h(\gamma) = 
\langle \Psi^{\text{UHF}}(0) | H | \Psi^{\text{UHF}}(\gamma) \rangle$
and
$n(\gamma) = 
\langle \Psi^{\text{UHF}}(0) | \Psi^{\text{UHF}}(\gamma) \rangle,$
where $\Psi^{\text{UHF}}(\gamma)$ is the original UHF determinant rotated by an
azimuthal angle $\gamma$ and $H$ is the many body Hamiltonian (including
the vector potential, external confinement, and Coulomb two-body repulsion).
We note that the UHF energies are simply given by $E_{\text{UHF}} = h(0)/n(0)$.

Compared to an EXD calculation, the CPU time for 
calculating the projected energies [see Eq.\ (\ref{eproj})] increases much 
slower as a function of $L$ and $N$. In addition, the computational efficiency
of Eq.\ (\ref{eproj}) is greatly enhanced by the knowledge \cite{kai}
of the analytical forms of the (in general complex) matrix elements of the 
single-particle and two-body components of $H$ between the displaced Gaussians.

\begin{figure}[t]
\centering\includegraphics[width=7.8cm]{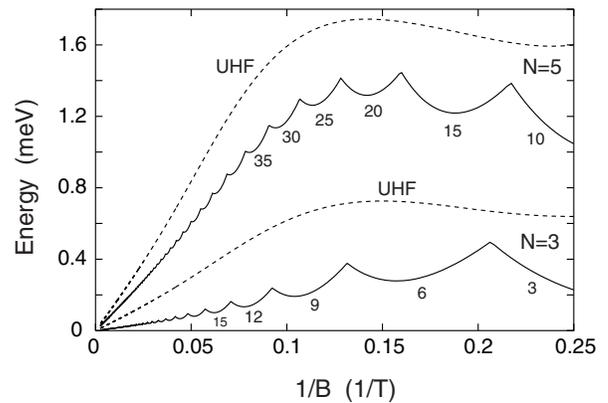}
\caption{
PRJ (solid lines) and UHF (dashed lines) total energies, $E-N \hbar \Omega$,
for the fully spin-polarized ground states of $N=3$ and $N=5$ electrons,
as a function of $1/B$. The range covered is 4 T $ \leq B \leq 400$ T, and the 
zero of energy corresponds to $E^{\text{st}}_{\text{cl}}= 14.415$ meV for
$N=3$ and to $E^{\text{st}}_{\text{cl}}=42.873$ meV for $N=5$.
The integers below the troughs of the PRJ curves indicate the corresponding 
magic angular momenta. The parameters are as in Fig.\ 1.
}
\end{figure}

To test the accuracy of our method, we display in Fig.\ 1 the total
energies $E-2 \hbar \Omega$ as a function of the magnetic field for the
triplet state of $N=2$ electrons and for all three levels of calculations, 
i.e., the UHF, the subsequent projection [PRJ, see 
Eq.\ (\ref{eproj})], and the exact result \cite{yl4}. We observe that the UHF 
energies fare poorly compared to the PRJ ones; 
note that for $B > 7$ T, the PRJ and exact results are practically the same.
We further note that 
the UHF energies do not reproduce the characteristic
oscillations of the exact ones; these oscillations originate from the fact that
only states with magic angular momenta can become ground states (in the case 
of the triplet state for $N=2$, the magic angular momenta are $L=2m+1$, 
$m=0,1,2,...$).

{\it Ground states at high magnetic field\/}.
Fig.\ 2 displays the ground state energies $E-N \hbar \Omega$ for $N=3$ and 
$N=5$ electrons as a function of $1/B$. Since our method allows us 
to reach magnetic field values as high as $B=400$ T, we conclude 
unequivocally that, as $B \rightarrow \infty$, both the PRJ and UHF energies 
approach the limiting value corresponding to the classical energy of $N$ 
point-like electrons at equilibrium [in the (0,N) configuration for $N=2-5$] 
inside an external parabolic confinement of frequency $\omega_0$, i.e.,  
\begin{equation}
E^{\text{st}}_{\text{cl}}(N)=
(3/8) (2 R_W)^{2/3} N S_N^{2/3} \hbar \omega_0,
\label{eclst}
\end{equation}
with $S_N= \sum_{j=2}^{N} \left( \sin[(j-1)\pi /N] \right)^{-1}$ \cite{note8}.
Except for the $B \rightarrow \infty$ limit, the UHF energies are higher in 
value than the projected ones.

The projected ground states have magic
angular momenta that vary (from one trough to the other) by steps of $N$ units
($N=3,4,5$) in accordance with the trend found for a much smaller range of 
angular momenta in earlier EXD studies \cite{mak,yang}. With our projection
method, we were able to reach remarkably high values of $L=369,690$, and 1100 
for $N=3,4$, and 5, respectively. The fact that  
$E_{\text{PRJ}}-N \hbar \Omega$ and $E_{\text{exact}}-
N \hbar \Omega$ (see Ref.\ \cite{yngl}) tend to $E^{\text{st}}_{\text{cl}}(N)$
for $B \rightarrow \infty$ is a significant finding that has not appeared in 
the FQHE literature \cite{jai,yang}; 
instead it has been implied that $E-N \hbar \Omega \rightarrow 0$ as 
$B \rightarrow \infty$. In any case, it is doubtful that the composite-fermion
approach can conform to this rigorous limit \cite{yngl}; indeed the precise 
value of $E^{\text{st}}_{\text{cl}}(N)$
depends crucially on the long-range character of the Coulomb force acting
between localized classical particles, and thus it is a result
innate to the WM picture.

\begin{figure}[t]
\centering\includegraphics[width=7.5cm]{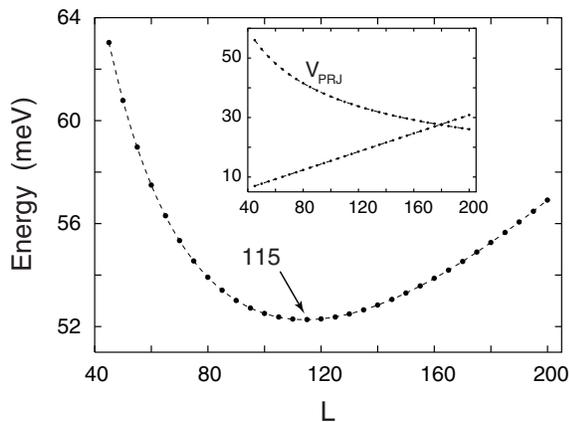}
\caption{
Projected total energies, $E_{\text{PRJ}}-5\hbar \Omega$, for the yrast 
rotational band (states with the lowest energy at magic angular momenta) of 
$N=5$ fully-polarized electrons at $B=60$ T, as a function of the angular 
momenta $L$. The ground state has angular momentum $L=115$. 
The choice of parameters is: $m^*=0.067m_e$, $\hbar \omega_0 = 
4$ meV, and $\kappa=12.9$.
Inset: The two contributions $E^c_{\text{PRJ}}(L)-5 \hbar \Omega$ and 
$V_{\text{PRJ}}(L)$ associated with the confinement (lower curve) and the 
Coulomb interaction (upper curve), respectively.
}
\end{figure}
Furthermore, this limit points out to interesting experimental
ramifications. Namely, at high $B$, the addition energies 
(expressed by the finite second difference $\Delta_2 E (N)$),
while not conforming to the capacitance$-$Coulomb blockade model \cite{yang}, 
will tend to a constant value 
$\Delta^{\text{st}}_{2,\text{cl}} E (N) = E^{\text{st}}_{\text{cl}}(N+1) + 
E^{\text{st}}_{\text{cl}}(N-1) - 2E^{\text{st}}_{\text{cl}}(N)$ (e.g., 
3.446 meV for $N=4$ and the parameters listed in the caption of Fig.\ 2); 
this constant value maybe detected experimentally.

{\it The yrast rotational band at high B\/}.
The {\it yrast band\/} consists of the lowest energy states (for a given $B$) 
having magic angular momenta (the ground state is a member of
this band and the remaining states are excited ones). The yrast band for
$N=5$ calculated at $B=60$ T is displayed in Fig.\ 3.

The energy of the yrast states is composed \cite{jai} of two 
contributions; namely, $E(L)=E^c(L) + V(L)$, where $E^c(L)$ is the confinement 
energy associated with the external potential (including the kinetic energy), 
and $V(L)$ is the interaction energy due to the Coulomb force. For high $B$, 
it is well known \cite{jai,yang} that the confinement energy varies linearly 
with $L$; in particular, $E^c(L) \rightarrow \hbar (\Omega - \omega_c/2)L + N
\hbar \Omega$.

We have calculated the projected values [see Eq.\ (\ref{eproj})] for these two 
components, namely, $E^c_{\text{PRJ}}(L)$ and $V_{\text{PRJ}}(L)$, for 
$N=2-6$ and a variety of $B$ values and QD parameters. Our ability to
calculate efficiently high $L$ values allowed us to confirm numerically the 
linear $L$-dependence of the confinement energy $E^c(L)$, and also the precise
proportionality coefficient $\hbar (\Omega - \omega_c/2)$.  
Moreover, rather unexpectedly, we also find that the interaction 
energy behaves asymptotically as $L^{-1/2}$, i.e., 
$V(L) \rightarrow C_V/L^{1/2}$, where $C_V$ is a constant given below.

\begin{table}[t]
\caption{\label{vl} 
Projected interaction energies $V_{\text{PRJ}}(L)$ associated with yrast 
states at two values of $B=60$ T and $B \rightarrow \infty$ for $N=5$ 
[the (0,5) configuration] and $N=6$ [the (1,5) configuration], respectively. 
$f \equiv V_{\text{PRJ}}(L) L^{1/2}/C_V$, where
$C_V=15.388$ for $N=5$ and $C_V=18.787$ for $N=6$.
The fractional fillings are calculated through $\nu=N(N-1)/(2L)$.
Energies in units of $e^2/\kappa \lambda$ and $e^2/\kappa l_B$ for
$B=60$ T and $B \rightarrow \infty$, respectively. At $B \rightarrow \infty$, 
$V(L)$ is independent of $\hbar \omega_0$. At large finite $B$, $V(L)$ 
depends on $\hbar \omega_0$ through $\lambda$.
}
\begin{ruledtabular}
\begin{tabular}{ccc|ccc}
 $N=5$ & $B=60$ T  &    & $N=6$ & $B \rightarrow \infty$ &   \\ \hline
 $L(\nu)$ & $V_{\text{PRJ}}$ & $f$ & $L(\nu)$ & $V_{\text{PRJ}} $ &$f$\\ \hline
 40(1/4)    & 2.5000  & 1.02749 & 75(1/5)   & 2.2196 & 1.02317 \\
 50(1/5)  & 2.2216  & 1.02083 & 135(1/9)  & 1.6361 & 1.01186 \\
 60(1/6)  & 2.0198  & 1.01669 & 195(1/13) & 1.3561 & 1.00799 \\
 80(1/8)  & 1.7409  & 1.01189 & 255(1/17) & 1.1836 & 1.00603 \\
100(1/10) & 1.5531  & 1.00925 & 315(1/21) & 1.0636 & 1.00484 \\
120(1/12) & 1.4154  & 1.00757 & 435(1/29) & 0.9039 & 1.00347 \\
140(1/14) & 1.3089  & 1.00641 & 495(1/33) & 0.8470 & 1.00304 \\
160(1/16) & 1.2233  & 1.00556 & 555(1/37) & 0.7996 & 1.00271 \\
180(1/18) & 1.1526  & 1.00491 & 615(1/41) & 0.7594 & 1.00244 \\
190(1/19) & 1.1216  & 1.00464 & 675(1/45) & 0.7247 & 1.00222 \\
\end{tabular}
\end{ruledtabular}
\end{table}

TABLE I lists $V_{\text{PRJ}}(L)$ for $N=5$ and $B=60$ T, as well as for 
$N=6$ [the (1,5) configuration] and $B \rightarrow \infty$; the latter limit 
can be easily taken by setting \cite{yl1} $\lambda = l_B \sqrt{2}$ in Eq 
(\ref{uhfo}), which restricts the electrons to the lowest Landau level.
From TABLE I, one can see that the quantity 
$f \equiv V_{\text{PRJ}}(L) L^{1/2}/C_V$ approaches unity as the angular
momentum intreases, where 
$C_V = K^{3/2} (S_K/4 + \delta_{K,N-1})\;e^2/(\kappa \lambda)$, with 
$K=N$ or $N-1$ for WM's in the $(0,N)$ or $(1,N-1)$ configurations, 
respectively.

The $C_V/L^{1/2}$ asymptotic dependence of the interaction energy can be 
readily associated with a model \cite{mak} of a classical {\it floppy 
molecule\/} rotating inside a parabolic confinement characterized by a 
frequency $\Omega$ \cite{note2}. Indeed, the energy of such a molecule 
is given by
\begin{eqnarray}
E^{\text{rot}}_{\text{cl}}(K)=&&
\hbar^2 L^2/(2{\cal J}(a)) + {\cal J}(a) \Omega^2/2 - \hbar \omega_c L /2 
\nonumber \\
&& + K (S_K/4 + \delta_{K,N-1})e^2/(\kappa a),
\label{eclrot}
\end{eqnarray} 
where ${\cal J}(a) = K m^* a^2$ is the moment of inertia of $K$ point-like 
electrons located at the vertices of a regular polygon of radius $a$. The 
second term is the potential energy due to the confinement, and the last term 
is the classical Coulomb-repulsion electrostatic energy. At given $B$,
the radius of this floppy molecule varies with $L$, and for large angular 
momentum (and/or high $B$) it is given by $a \approx \lambda \sqrt{L/K}$
[$a \rightarrow \sqrt{2 \hbar c L /(eBK)}$ for $B \rightarrow \infty$].  
Substitution into Eq.\ (\ref{eclrot}) yields the aforementioned linear and
$1/L^{1/2}$ contributions, i.e.,
\begin{equation}
E^{\text{rot}}_{\text{cl}} (K) \approx 
\hbar (\Omega - \omega_c/2) L + C_V /L^{1/2}.
\label{eclrot2}
\end{equation}

A semiclassical approximation $L^{\text{cl}}_{\text{gs}}$ of the ground-state 
angular momentum minimizes Eq.\ (\ref{eclrot2}), yielding
$L^{\text{cl}}_{\text{gs}} \propto B$. Consequently, the ground-state radius
$a_{\text{gs}} \rightarrow r_0$ as $B \rightarrow \infty$, where 
$r_0 =l_0 R_W^{1/3} (S_K/4 + \delta_{K,N-1})^{1/3}$ is the radius of the 
{\it static\/} classical Wigner molecule. Note that, in the ground state as 
$B$ increases, the WM rotates faster in order to maintain the requisite value 
of $a_{\text{gs}}=r_0$.

We note that this inverse-square-root-of-$L$ law at constant $B$ for the 
Coulomb interaction energy has also been overlooked in the FQHE literature 
\cite{jai,yang}. This law provides a nontrivial test in that it lends further
support to the RWM picture; especially since it also applies in the 
$B \rightarrow \infty$ limit (see right part of TABLE I), which restricts
the electrons to the lowest Landau level (FQHE regime). We note that high 
$L$'s correspond to lower fractional fillings, experimentally achieveable in 
high-mobility samples.  

\begin{table}[t]
\caption{\label{b0}
Projected total energies $E_{\text{PRJ}}(L)$ at $B=0$ and $R_W=200$ associated
with yrast states for $N=5$ electrons [(0,5) configuration].
$ \protect\tilde{f} \equiv C_R/C_R^{\text{cl}}$ (see text).
Energies in units of $\hbar \omega_0$.
}
\begin{ruledtabular}
\begin{tabular}{ccc|ccc}
$L$ & $E_{\text{PRJ}}$ & $\protect\tilde{f}$ & 
$L$ & $E_{\text{PRJ}} $ & $\protect\tilde{f}$\\ \hline
  0  & 323.3070   &        & 25  & 324.7657 & 0.988 \\
  5  & 323.3656   & 0.992  & 30  & 325.4033 & 0.986 \\
 10  & 323.5414   & 0.992  & 35  & 326.1537 & 0.983 \\
 15  & 323.8338   & 0.991  & 40  & 327.0153 & 0.981 \\
 20  & 324.2422   & 0.989  & 45  & 327.9866 & 0.978 \\                   
\end{tabular}
\end{ruledtabular}
\end{table}

{\it The rigid rotor at zero magnetic field.}
At $B=0$, we found it advantageous to allow the width $\lambda$ and the
positions $Z_j$'s of the displaced Gaussians 
to vary in order to minimize (for each $L$) the projected 
energy [Eq.\ (\ref{eproj})]. For large $R_W$'s and for $N=2-5$ electrons, we 
find that the energies of the states in the yrast band can be approximated by
\begin{equation}
E_{\text{PRJ}}(L) \approx E_{\text{PRJ}}(0) + C_R L^2,
\label{erigid}
\end{equation}
where the rigid-rotor coefficient $C_R$ is essentially a constant whose value 
is very close (see the $\widetilde{f}$ values) to the classical one, 
corresponding to point charges in their $(0,N)$ equilibrium configuration 
inside a parabolic confinement of frequency $\omega_0$; i.e., 
$C_R \approx C_R^{\text{cl}} = \hbar^2 /(2 {\cal J}(r_0))$.

In Table \ref{b0}, we list calculated 
$E_{\text{PRJ}}(L)$ values for $N=5$ when $R_W=200$. 
The fact that $\widetilde{f} \equiv C_R/C_R^{\text{cl}} \approx 1$ for 
$L \leq 45$ illustrates that the RWM behaves as a quasiclassical rigid rotor. 
For smaller values of $R_W$, the rigidity of the RWM is progressively 
reduced \cite{yl4}. We note that $E_{\text{PRJ}} (0) - N \hbar \omega_0$ is 
very close to the classical electrostatic value given by Eq.\ (\ref{eclst}).

{\it Conclusions.\/}
Using the method of restoration of broken circular symmetry, we have
shown that the RWM's in QD's exhibit characteristic properties as a result of
their collective rotation: (I) in high $B$, they behave like 
{\it floppy rotors\/} and the energies of the magic states in the lowest 
rotational (yrast) band, associated with magic angular momenta, obey 
$a L + b/L^{1/2}$. In addition, the ground-state energies (referenced to the 
kinetic energy in the lowest Landau level) approach the limit of the 
electrostatic energy of $N$ point charges in their classical equilibrium 
molecular configuration; (II) at $B=0$ and strong interelectron repulsion, 
the RWM's behave like quasiclassical {\it rigid rotors\/} whose yrast band 
exhibits an $L^2$-dependence. The rotating-Wigner-molecule picture unifies the 
description of the various physical regimes of strongly correlated electrons 
in 2D QD's. This unification is achieved through the computationally powerful 
method of restoration of broken symmetries via projection techniques.

This research is supported by the U.S. D.O.E. (Grant No. FG05-86ER45234).


\begin{thebibliography}{99}
\bibitem{note8}
For the same parameters
as those in Fig.\ 2, we have also calculated the ground-state energies 
$E_{\text{PRJ}}-4 \hbar \Omega$ for $N=4$ electrons in the range 4 T $ \leq B
\leq 400$ T, and found that they converge to $E^{\text{st}}_{\text{cl}}(4)=
26.921$ meV.
\end{thebibliography}
\end{document}